# Photoinduced features of energy band gap in quaternary $Cu_2CdGeS_4$ crystals

M G Brik[a,1], I V Kityk[b], O V Parasyuk[c], G L Myronchuk[d]

[a] Institute of Physics, University of Tartu, Riia 142, Tartu 51014, Estonia
[b] Electrical Engineering Department, Czestochowa University of Technology, Armii Krajowej 17, PL-42-201 Czestochowa, Poland
[c] Chemical Department, Eastern European University, Voli 6, Lutsk, Ukraine
[d] Physical Department, Eastern European University, Voli 6, Lutsk, Ukraine

**Abstract**

Quaternary chalcogenide crystal $Cu_2CdGeS_4$ was studied both experimentally and theoretically in the present paper. Investigations of polarized fundamental absorption spectra demonstrated a high sensitivity to external light illumination. The photoinduced changes were studied using the cw 532 nm green laser with energy density about 0.4 J/cm$^2$. The spectral maximum of the photoinduced anisotropy was observed at spectral energies equal to about 1.4 eV (energy gap equal to about 1.85 eV) corresponding to maximal density of the intrinsic defect levels. Spectroscopic measurements were performed for polarized and un-polarized photoinducing laser light to separate contribution of the intrinsic defect states from the pure states of the valence and conduction bands. To understand the origin of the observed photoinduced absorption near the fundamental edge, the benchmark first-principles calculations of the structural, electronic, optical and elastic properties of $Cu_2CdGeS_4$ were performed by the general gradient approximation (GGA) and local density approximation (LDA) methods. The calculated dielectric function and optical absorption spectra exhibit some anisotropic behavior (shift of the absorption maxima in different polarizations) within the 0.15…0.20 eV energy range not only near the absorption edge; the optical anisotropy was also found for the deeper inter-band transition spectral range. Peculiar features of chemical bonds in $Cu_2CdGeS_4$ were revealed by studying the electron density distribution. Possible intrinsic defects are shown to affect considerably the optical absorption spectra. Pressure effects on the structural and electronic properties were modeled by optimizing the crystal structure and calculating all relevant properties at elevated hydrostatic pressure. The first estimations of the bulk modulus (69 GPa (GGA) or 91 GPa (LDA)) and its pressure derivative for $Cu_2CdGeS_4$ also are reported.

---

[1] Corresponding author: brik@fi.tartu.ee

1. Introduction

Quaternary chalcogenide compounds like $Cu_2CdGeS_4$ present an extreme interest both from the fundamental and applied points of view due to their possible applications as photovoltaic and optoelectronic materials [1,2]. Besides, it is also important to see how the fundamental parameters of the energy gap are varied when moving from binary to ternary and further to quaternary compounds. One of specific features of these ternary/quaternary compounds is a high number of intrinsic defects [3], which favors very interesting photoinduced polarizabilities that cannot be explained only within a framework of perfect crystalline long-range ordered band structures. Thus, in the present work we will compare the experimental data on the photoinduced changes caused by the polarized light with the density functional theory (DFT) calculations in order to resolve the role of the local intrinsic disorder in the effects observed.

The studied $Cu_2CdGeS_4$ single crystals have a general chemical formula $A^I_2B^{II}C^{IV}D^{VI}_4$ and belong to the class of the diamond-like semiconducting crystals [4]. The structural, thermodynamic and optical properties of $Cu_2$-II-IV-$VI_4$ quaternary compounds were studied by H. Matsushita *et al* [5] and M. Himmrich *et al* [6]. The chosen crystal has an advantage with respect to others because it possesses simultaneously coordination polyhedra typical for the chalcogenide crystals as well as Cu ions which favor substantial light photopolariztion [7] due to the d-p charge transfer. The acentric structure of the title crystal may also indicate a possibility of their applications as materials for the second harmonic generation in the infrared spectral range. Many representatives of this family of compounds also seem to have enormous potential as photogalvanic materials [8, 9, 10].

However, their further application is restrained by absence of reliable band structure parameters closely related to the principal band energy gap, which could give an additional important information concerning the effective masses, space contour plots of the electron density, origin of chemical bonds, carrier mobility etc [11]. Additionally, for the quaternary chalcogenide crystals a number of the intrinsic defects [12, 13, 14, 15] is almost one order higher than for other chalcogenides. As a consequence, the electronic sub-bands formed by these defects should be taken into account, even if it is quite a laborious task to tackle by one-electron band structure calculations. The existence of such band tails formed by intrinsic trapping levels may significantly change the materials' optoelectronic properties [16] especially if the surrounding conditions (temperature, pressure, etc) are varied. So, to assess their potential for further applications, it is necessary not only to perform the standard band structure calculations within the DFT approach [17], but to verify the obtained calculated results at least in the spectral range corresponding to the band gap and its vicinity. Very often the scissor operator [18] and different exchange correlation potentials [19] should be used in the DFT calculations for a reasonable description of the studied compounds.

One of specific features of the quaternary chalcogenide compounds is a coexistence of different crystallochemical polyhedral clusters with chemical bonds of different origin and types [20]. Therefore, information on the origin of the chemical bonds within different structural fragments is a vital condition for successful engineering of new materials with desired physical properties. As far as $Cu_2CdGeS_4$ is concerned, the previous studies have shown that the principal role for the optoelectronic studies is played by the Ge-S tetrahedra [21]. At the same time, influence of both Cu and Cd cations also is important in $Cu_2CdGeS_4$. Usually there is no specific rule how to study an influence of different kinds of anions on physical properties of crystals. For the case of a large number of the intrinsic defects in the crystal lattice, the band structure features should crucially depend on the number of defects and imperfections [22]. This factor is usually neglected, which leads to difficulties when reproducing the necessary experimental results.

Another principal factor for the chalcogenide materials is a high degree of the electron-phonon interaction [23] including anharmonic effects [24]. The latter factor can substantially broaden the corresponding spectral bands [25] and even could lead to their spectral shifts or splitting [26]. This factor also plays some role in the nonlinear optical features [27], especially in the photoinduced nonlinear optical effects [28]. The only criterion to evaluate the strengths of the intrinsic defects as well as phonon contribution is to compare the DFT calculations with the experimental data at least near the absorption edge using polarized spectroscopy.

However, it is very difficult to prepare high-quality crystals of quaternary chalcogenides as compared with ternary semiconductors. Thus, it is necessary to establish structural properties to obtain single crystals with suitable size, compositional homogeneity and the reliable databases to reproduce the physical properties.

As a consequence, the present work will be devoted to the studies of the band structure of the title compound using the DFT calculations and comparison of the calculated results with the experimental polarized spectroscopy data.

The $Cu_2CdGeS_4$ compound, which is crystallized in the structural wurtzite-like type (space group *Pmn2$_1$*) [29], was previously just studied with respect to temperature-dependent electrical conductivity, Hall mobility, thermoelectric power, and the absorption coefficient, as reported in Refs. [21, 30].

The main goal of this article is to explore the photoinduced changes near the energy band gap and to clarify the role of the intrinsic defects for the observed phenomena. Previous studies on the photoinduced effects in the chalcogenides [22, 28] were never focused on the changes of the band structure obtained from the DFT calculations, including the effects of the external hydrostatic pressure on the electronic and optical properties of the studied material. The latter may highlight the role of the different kinds of voids on the energy band gap behavior [31] and give an information about the possible contribution of particular chemical bonds to the effects observed. One can expect that such combination of the experimental and DFT-based methods would allow for clarifying existed discrepancies between the calculated and experimental band gap values as well as evaluating contribution of the local disorder into the observed effects.

The structure of the paper is as follows: in the next section the crystal growth details are presented; then we proceed with the description of the method of calculations, calculated results and their relation to the experimental data, which were also obtained by us. After discussion of the obtained theoretical and experimental results, we conclude the paper with a short summary.

**2. Crystal growth**

The bulk crystals of $Cu_2CdGeS_4$ were obtained by Bridgman-Stockbarger technique. Initially by applying horizontal gradient crystallization the single crystals in the shape of a parallelepiped of a size ~2x2x8–12 mm were obtained. The crystals were oriented along the principal crystallographic axis.

The process of melting and recrystallization annealing was lasted during 100 hours. The growth rate was 0.1-0.15 mm/h, and the temperature gradient at the "crystal-melt" boundary was equal to 2-3 K/mm. After solidification of the melt, the crystal was annealed during 100 h at 870 K, and then slowly cooled down to ambient temperature.

$Cu_2CdGeS_4$ crystal structure was refined by X-ray powder on a DRON 4-13 using Cu $K_\alpha$ rays. Diffraction pattern analysis was indexed in the orthorhombic structure in the model proposed in Ref. [29]. Inter-atomic distance and valence angles in the $Cu_2CdGeS_4$ structure are all in good agreement with those for similar compounds [32, 33, 34] and are close to the results obtained for a single crystal [29].

**3. Method of calculations**

The CASTEP module [35] of Materials Studio package was used to calculate the structural, electronic, optical and elastic properties of $Cu_2CdGeS_4$ single crystal. The initial structural data for this compound were taken from Ref. [36]. One unit cell of $Cu_2CdGeS_4$ is depicted in Fig. 1, whereas Fig. 2 shows the corresponding Brillouin zone (BZ) for this structure and the principal high symmetry point identification. Each ion in this structure is 4-fold coordinated. The generalized gradient approximation (GGA) with the Perdew-Burke-Ernzerhof [37] and the local density approximation (LDA) with the Ceperley-Alder-Perdew-Zunger (CA-PZ) functional [38, 39] were used to treat the exchange-correlation effects. The Monkhorst-Pack **k**-points grid was chosen as 4×4×4 grid. The cut-off energy, which determines the size of the plane-wave basis set, was 320 eV. The convergence criteria were as follows: $5×10^{-6}$ eV/atom for energy, 0.01 eV/Å for maximal force, 0.02 GPa for maximal stress and $5×10^{-4}$ Å for maximal displacement. The electronic configurations were as follows: $3d^{10}4s^1$ for Cu, $4d^{10}5s^2$ for Cd, $4s^24p^2$ for Ge, and $3s^23p^4$ for S, and the ultrasoft pseudopotentials were used for all chemical elements.

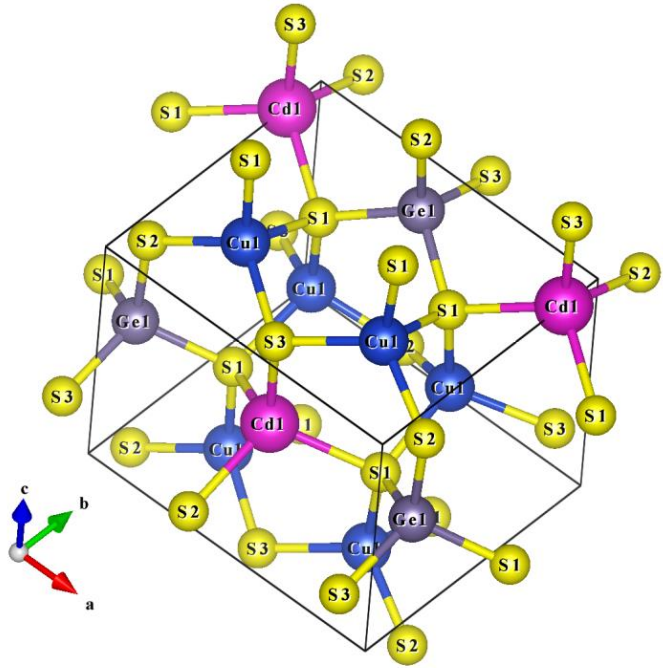

Fig. 1. A unit cell of $Cu_2CdGeS_4$. Drawn by VESTA [40].

## 4. Results of calculations: structural, electronic and optical properties at ambient pressure

Table 1 contains a summary of the experimental and optimized structural data for $Cu_2CdGeS_4$ single crystal. Comparison of the calculated and experimental results yields a good agreement (the maximal relative difference between the experimental and optimized lattice parameters is 0.9 % in the case of the GGA calculations, and 2.4 % in the case of the LDA calculations), which makes a firm basis for reliability of the subsequent analysis of the electronic, optical and elastic properties of this material.

Table 1. Summary of the structural data for $Cu_2CdGeS_4$.

| | Exp. [36] | | | Calc. (this work) | | | | | |
|---|---|---|---|---|---|---|---|---|---|
| | | | | GGA | | | LDA | | |
| $a$, Å | 7.7024 | | | 7.7705 | | | 7.5281 | | |
| $b$, Å | 6.5486 | | | 6.6067 | | | 6.3915 | | |
| $c$, Å | 6.2928 | | | 6.3499 | | | 6.1653 | | |
| Atomic coordinates | | | | | | | | | |
| | x/a | y/b | z/c | x/a | y/b | z/c | x/a | y/b | z/c |
| Cu | 0.2504 | 0.3270 | 0 | 0.2510 | 0.3283 | 0.0022 | 0.2513 | 0.3276 | 0.0028 |
| Cd | 0 | 0.8411 | 0.9971 | 0 | 0.8436 | 1.0005 | 0 | 0.8439 | 1.0010 |
| Ge | 0 | 0.1786 | 0.498 | 0 | 0.1781 | 0.498 | 0 | 0.1781 | 0.4979 |
| S1 | 0.2335 | 0.3420 | 0.3688 | 0.2304 | 0.3463 | 0.3647 | 0.2311 | 0.3477 | 0.3637 |
| S2 | 0 | 0.2068 | 0.864 | 0 | 0.2065 | 0.8559 | 0 | 0.2082 | 0.8562 |
| S3 | 0 | 0.8480 | 0.398 | 0 | 0.8444 | 0.4065 | 0 | 0.8429 | 0.4065 |

According to Ref. [36], the band gap of $Cu_2CdGeS_4$ is 2.05 eV. The crystals studied here have a smaller experimental energy gap (about 1.85 eV), which reflects underestimation of the band gap due to the existence of the mentioned intrinsic defect trapping states and indicates dependence of the presence and concentration of these defects on the preparation conditions. The calculated energy gap values were 0.372 eV (GGA) and 0.682 (LDA). Both results are underestimated in comparison to the experimental result. This underestimation is a common feature of the DFT-based methods, and to overcome it, a scissor operator can be used, which simply shifts upward the conduction band to make the band gap matching the experimental data. In our case, the values of such a shift were 1.37 eV (LDA) and 1.68 eV (GGA). Fig. 3 presents both calculated band structures with taking into account the scissor operator. The band edge is of a direct type, since the maximum of the valence band and the minimum of the conduction band are both realized at the BZ center.

The observed discrepancies between various experimental data reflect influence of the intrinsic defect tails, which produce additional absorption near the absorption edge resulting in a difference in the experimental band gap estimations. This also may indicate that the crystal growth conditions may be responsible for these deviations. At the same time the phonon subsystem will lead to additional broadening of the absorption edge and corresponding spectral lines. On the theoretical level, the GGA and LDA approaches also produce somewhat different results due to a different level of approximations involved.

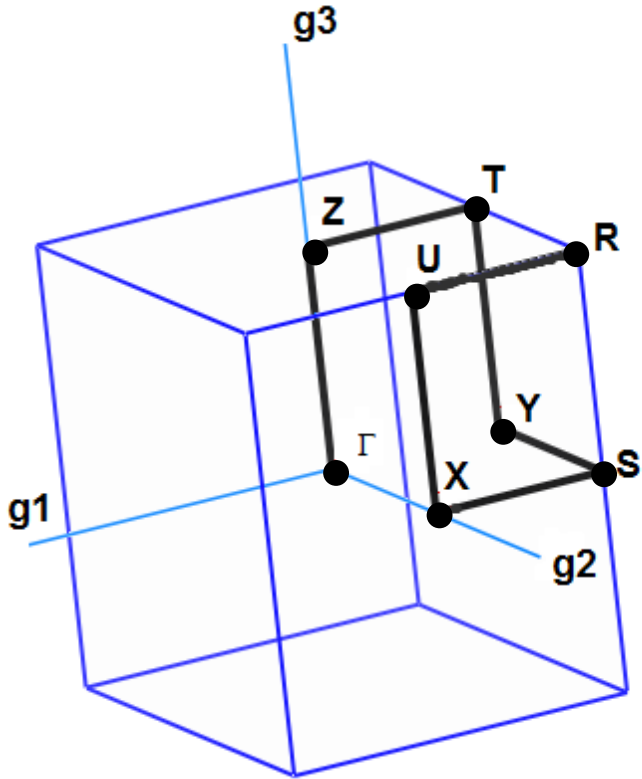

Fig. 2. The Brillouin zone of a unit cell for $Cu_2CdGeS_4$. The coordinates of the special points of the Brillouin zone are (in units of the reciprocal lattice vectors); Γ (0, 0, 0); Z (0, 0, 1/2), T(-1/2, 0, 1/2), Y(-1/2, 0, 0), S(-1/2, 1/2, 0), X(0, 1/2, 0), U(0, 1/2, 1/2), R(-1/2, 1/2, 1/2).

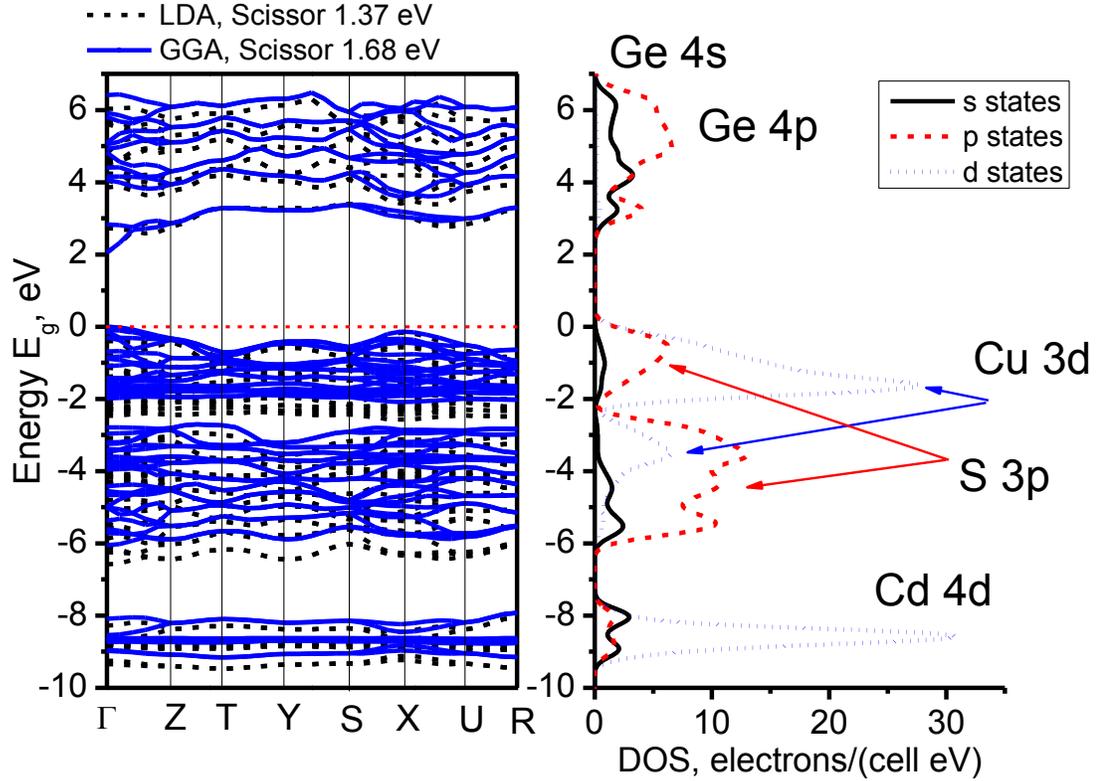

Fig. 3. Calculated band structure and density of states (DOS) of $Cu_2CdGeS_4$: LDA and GGA results. The values of the scissor operators are indicated. Only the GGA data were used in the DOS diagrams for the sake of brevity.

Following the results presented in Fig. 3 one can clearly see that the principal difference in the band dispersion in the **k**-space exists between the S-X-U BZ direction in the plane perpendicular to the crystallographic axis and in the Z-T-Y direction perpendicularly to the axis. Also in the Z-T-Y direction one can see a principal difference between the LDA and GGA results. The effective masses of the holes are the lowest near the top of the valence band, i.e. along the S-X-U and Γ-Z BZ directions. The topmost states of the valence band, as obtained in the GGA and LDA runs, are very close.

It is crucial that the top of the valence band is formed prevailingly by the relatively localized Cu 3d states and the bottom of the conduction band is originated from the delocalized Ge 4p,4s states, So the d-p charge transfer plays here the crucial role. As a consequence the intrinsic defect sub-bands will be partially overlapped with the bands formed by the above-mentioned orbitals. Fig. 3 also shows the partial density of states (DOS) diagrams, which allow to find the composition of the calculated electronic bands as follows: the conduction band (about 3 eV wide) is made basically of the Ge 4s, 4p states. The valence band, whose width is about 7 eV, has a complicated structure and is formed prevailingly by the Cu 3d, S 3p states, with a slight admixture of the Ge 4s, 4p states due to hybridization. The Cu 3d states exhibit two clearly pronounced spectral peaks, whose barycenters are situated at about -4.08 eV and -2.03 eV (indicated by the arrows in Fig. 3), which can be interpreted as the splitting of the 3d Cu states in a tetrahedral crystal field into the $e_g$ and $t_{2g}$ states, respectively. The latter are additionally disturbed by the Jahn-Teller effect (splitting due to existence of the phonon subsystem). The Cd 4d states form a sharp peak at about -8.7 eV, and the Ge 4s states are peaked at -8.2 eV and -9.2 eV. Finally, the lowest energetic bands are formed by the S 3s states with the peaks at about -14.6 eV and -13.4 eV (not shown in the figure).

It is clear that the Cu 3d-originated band is more flat in **k**-space and the conduction band is more delocalized. So the mobility of the holes will be substantially lower than that one of the electrons. This fact may play a significant role during the optically stimulated occupations of the trapping sub-bands. At the same time the S 3p sub-band is clearly separated by energy with respect to the top of valence band, which favors separation of the hole carriers by their mobility with respect to the energy. This is principally different with respect to other chalcogenides, where the top of the valence band was formed by the halides' p- states [41]. Such specific combination of the band structure parameters together with a presence of the mentioned intrinsic defects may cause very interesting optically induced effects [41], like photoinduced anisotropy [42], piezoelectricity [43] etc. Here it should be added that for the case of the corresponding glasses the band edge will not be so clear like for the crystals due to the Anderson disorder, which restrains their use for this reasons.

The macroscopic (obtained from experiments) optical properties of a solid usually are analyzed following the dispersion of optical functions [44]. The imaginary part Im($\varepsilon(\omega)$) of a dielectric function $\varepsilon(\omega)$ (directly related to the absorption spectrum of a solid) is calculated by numerical integrations in **k**-space of dipole matrix operator elements between the occupied states in the valence band and empty states in the conduction band:

$$\mathrm{Im}(\varepsilon(\omega)) = \frac{2e^2\pi}{\omega\varepsilon_0} \sum_{k,v,c} \left|\left\langle \Psi_k^c \left| \vec{u}\cdot\vec{r} \right| \Psi_k^v \right\rangle\right|^2 \delta\left(E_k^c - E_k^v - E\right), \qquad (1)$$

where $\vec{u}$ is the polarization vector of the incident electromagnetic field; $\vec{r}$ and $e$ are the electron's position vector and electric charge, respectively, $\Psi_k^c, \Psi_k^v$ are the wave functions of the conduction and valence bands at BZ **k** point, respectively; $E = \hbar\omega$ is the incident photon's energy; $\varepsilon_0$ is the vacuum dielectric permittivity. The summation in Eq. (1) is carried out over all band states originating from the occupied and empty bands, with the appropriate wave functions obtained in a numerical form after optimization of the initial crystal structure.

The real part Re($\varepsilon(\omega)$) of the dielectric function $\varepsilon$, which determines the dispersion properties and refractive index values, is estimated in the next step by using the Kramers-Kronig relation [45]:

$$\mathrm{Re}(\varepsilon(\omega)) = 1 + \frac{2}{\pi} \int_0^\infty \frac{\mathrm{Im}(\varepsilon(\omega'))\omega' d\omega'}{\omega'^2 - \omega^2} \qquad (2)$$

The calculated dispersions of real and imaginary parts for $Cu_2CdGeS_4$ are shown in Fig. 4.

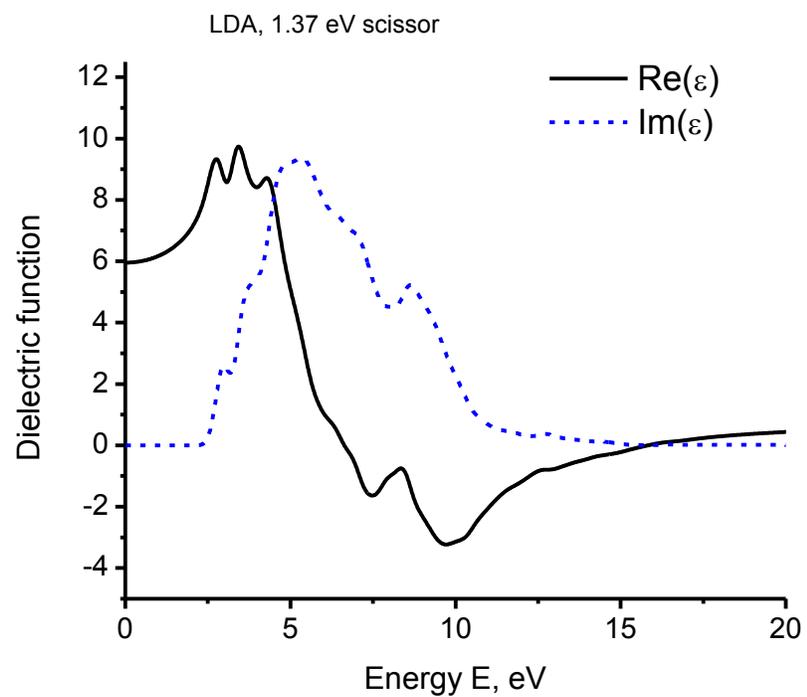
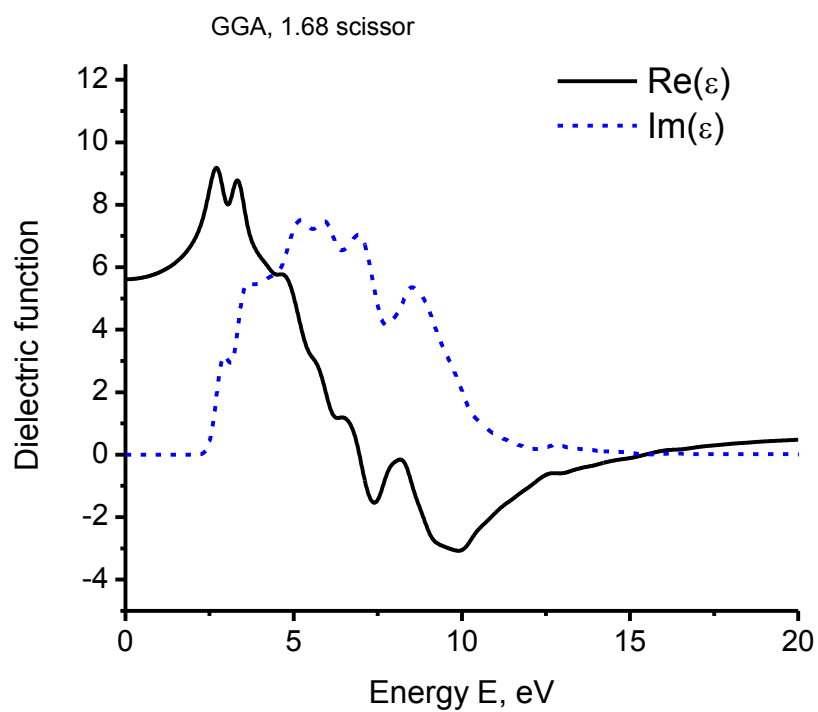

Fig. 4. Calculated dielectric functions dispersion for $Cu_2CdGeS_4$.

Taking the square root from the values of $\text{Re}(\varepsilon(\omega))$ in the zero limit energy, one can estimate the value of the refractive index, which is equal then to 2.45 (LDA) and 2.36 (GGA). Such discrepancies may be caused by different exchange-correlation screening potential used in the mentioned model. At the same time following Fig. 4 one can see that at about 5.5-6 eV the real part of the dielectric function changes its sign, which indicates an occurrence of the plasmonic resonances in this energy region.

**5. Experimental results and their relation to the calculated optical and electronic properties**

Fig. 5 shows the experimental polarized absorption spectra of $Cu_2CdGeS_4$ crystal near the fundamental absorption edge. The spectral resolution about 0.5 nm allows to separate spectrally the principal features of the absorption bands.

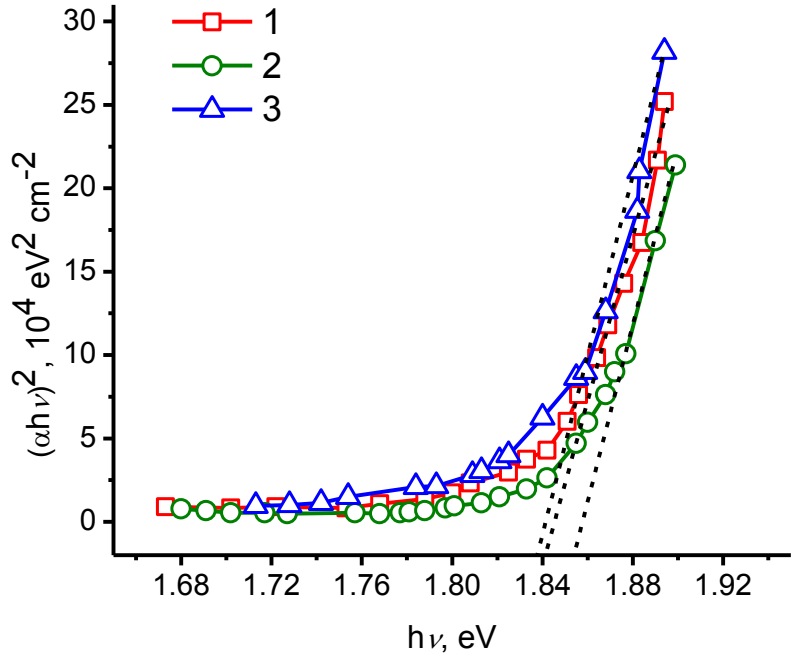

Fig. 5. Absorption edge for the $Cu_2CdGeS_4$ single crystal for different polarizations: 1 – non-polarized light; 2 - $\vec{E} \parallel c$; 3 - $\vec{E} \perp c$.

Following Fig. 5, one can clearly see that the value of the absorption anisotropy is almost stable in the whole considered spectral range, which indicates a fact that the origin of this anisotropy is mainly caused by anisotropy of the chemical bonds and corresponding inter-band transitions. Generally, the inter-band transitions for different light polarizations have some anisotropy of the transition dipole moment's magnitudes, which are closely related with the corresponding charge density distribution around ions in a crystal lattice. In the case of the considered crystal, this is caused by the space charge distribution of the chemical bonds coordinated by the S 3p states, which form the principal valence bands. The type of the absorption edge corresponds likely to the direct optical transitions [46], however there is an obvious some long-range tail responsible for the presence of the defect states in the band gap, which also give some contribution to the optical and photoinduced absorption. The degree of anisotropy between two absorption directions determined by the differences in the energy gaps (in the plane of the crystals, i.e. E||**c** and E$\perp$**c**) is about 0.3 eV. In most previous works such kind of anisotropy was explained by several photo-structural effects contributing to that phenomenon including photo-expansion, photo-darkening, and permanent self focusing [47]. However, in almost all earlier works there was no comparison offered with the charge density distribution obtained from the first principles band structure calculations. An additional factor to be mentioned may be a consequence of the disordered intrinsic defect, which (contrary to the inter-band transitions) are not sensitive to the optical transition rules in the dipole approximation [48]. However they may be more important during the photoinduced absorption due to the relatively higher polarizabilities. Following the previous works on the disordered chalcogenides, it was shown that they possess a large number of intrinsic defects [49], which under the influence of the external light below the energy gap can cause some photoinduced polarization. However, it is not clear how it may be related to their band structure origin not only near the absorption edge, but also with respect to deeper inter-band excitations. Generally the absorption edge studies present a sensitive tool, which reflects contribution of the deep inter-band oscillators [50], as well as the resonances near the absorption edge, like impurities, defects etc [51].

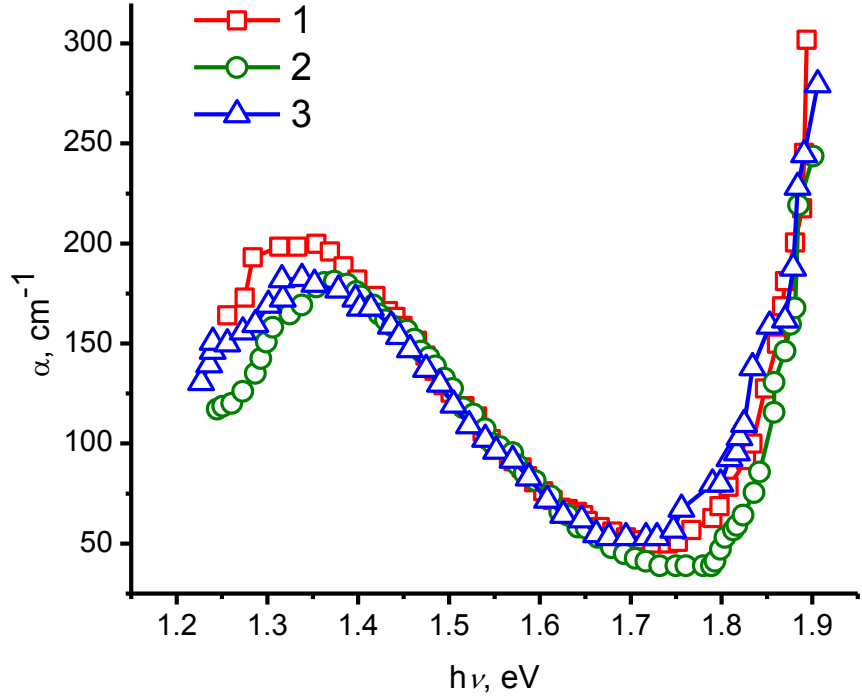

Fig. 6. Photoinduced changes of the absorption coefficients for the $Cu_2CdGeS_4$ during illumination by: 1 –un-polarized light; 2 - $\vec{E} \parallel \mathbf{c}$; 3 - $\vec{E} \perp \mathbf{c}$.

The previous studies have also shown that excitation of the levels below the energy gap may cause substantial changes in the absorption [22] due to re-occupation of such states as well as due to significant photopolarization of the intrinsic trapping levels. Additionally some role is played by electron-phonon broadening, which is extremely crucial for the chalcogenides [52].

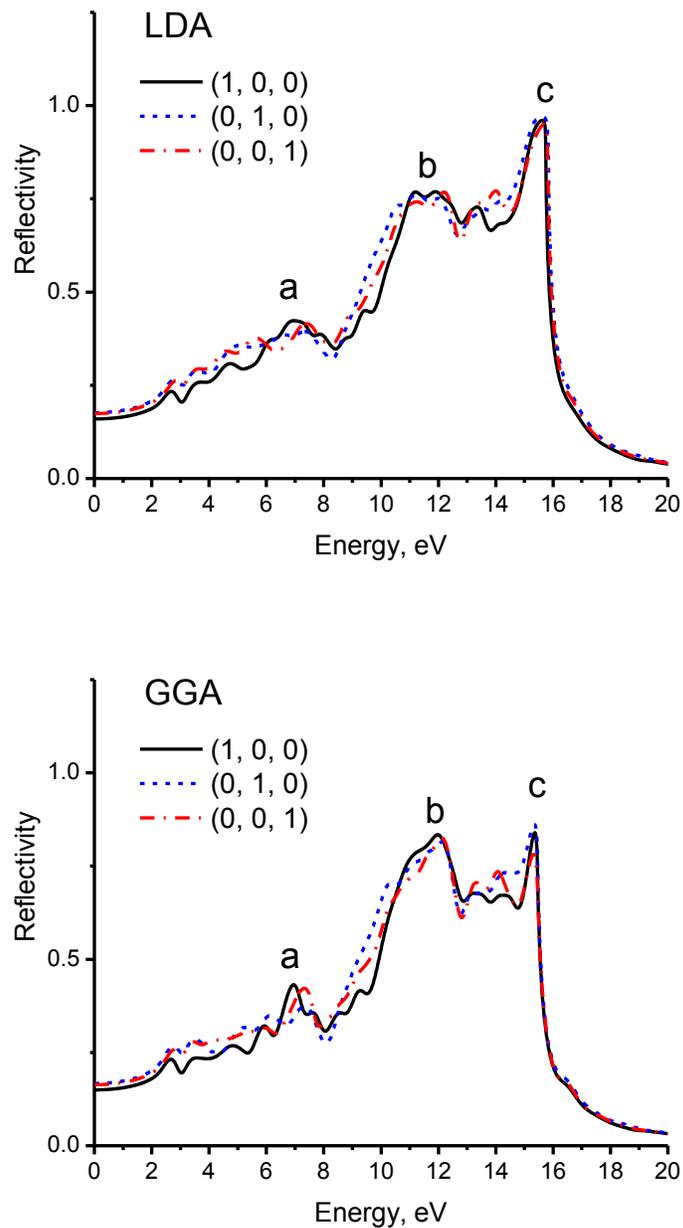

Fig. 7. Calculated polarized reflectivity spectra of $Cu_2CdGeS_4$.

Fig. 6 presents the photoinduced absorption caused by the 532 nm laser wavelength, which excites the localized trapping formed sub-bands. One can see that the maximal photoinduced anisotropy near the absorption edge is observed for the non-polarized light, which confirms the principal role of the trapping levels which are prevailingly isotropic for the external light. To explore the influence of crystalline long range ordering on deeper in energy contributions, we have performed *ab initio* simulation of the optical reflection spectra for the different light polarization (Fig. 7). Following the data presented in Fig. 7, one can see that maximal anisotropy in the corresponding spectra appeared for the *a* and *b* spectral peaks (whose maxima are shifted with respect to each other in different polarizations by 0.28…0.34 eV) and it may be neglected for deeper states indicated by *c* (where the such shift is about 0.03 eV only). It is interesting that the LDA approach gives more anisotropic states for the mentioned spectral peaks with respect to the GGA. We attribute such a difference to the fact that the LDA-optimized crystal parameters are smaller than the GGA ones. Therefore, in the LDA-optimized crystal lattice the atoms are situated closer to each other, they form stronger chemical bonds with a higher degree of hybridization of the atomic orbitals, which then leads to a more pronounced optical anisotropy.

The origin of such anisotropy is confirmed by non-uniformity of the charge density distribution for the two principal crystal planes (see Fig. 8, which shows the calculated cross-sections of the electron density difference in the space between the atoms in the $Cu_2CdGeS_4$ crystal lattice). One can see an obvious coexistence of the ionic and covalent types of chemical bonding. And the different plane sections confirm the anisotropy of the bond lengths in different directions. Particularly important is the anisotropy for the Cu and Ge atoms forming the principal band energy gaps (see Fig. 3). To understand the origin of the anisotropy we can compare the space charge density distribution for the Cu-S-Cu bonds formed prevailingly by the Cu 3d and S 3p states, respectively (top of Fig. 8) and the Cd-S-Cd charge density distribution in the ac direction (bottom of Fig. 8) direction. As a consequence the coordination of the S anions charge density distribution around them is different due to different degree of their hybridization with the Cu and Cd states.

The above presented reasons show that the role of anisotropy is not so crucial for the photoinduced changes like the presence of the randomly oriented intrinsic defect states for optical absorption.

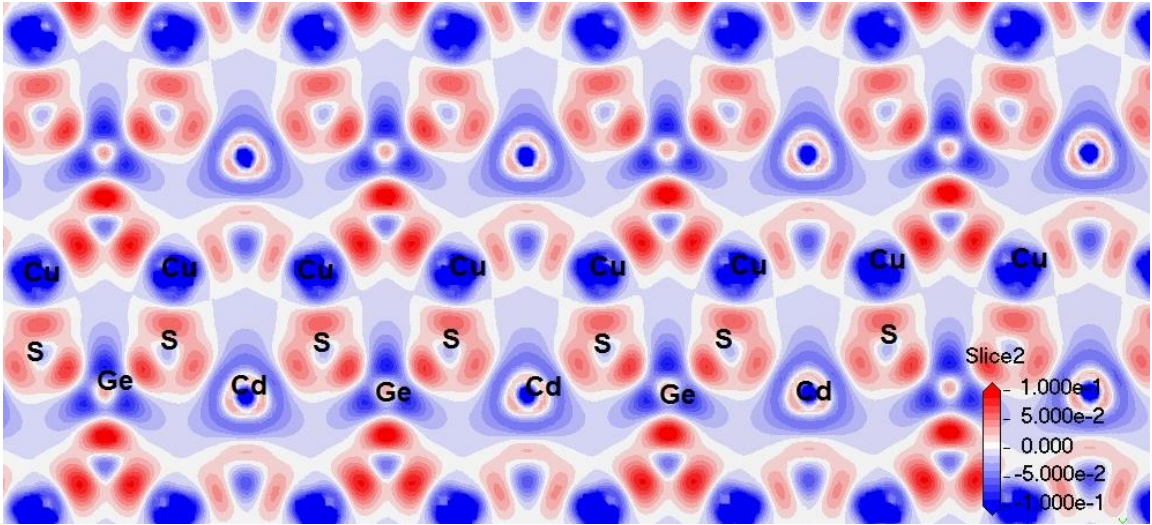

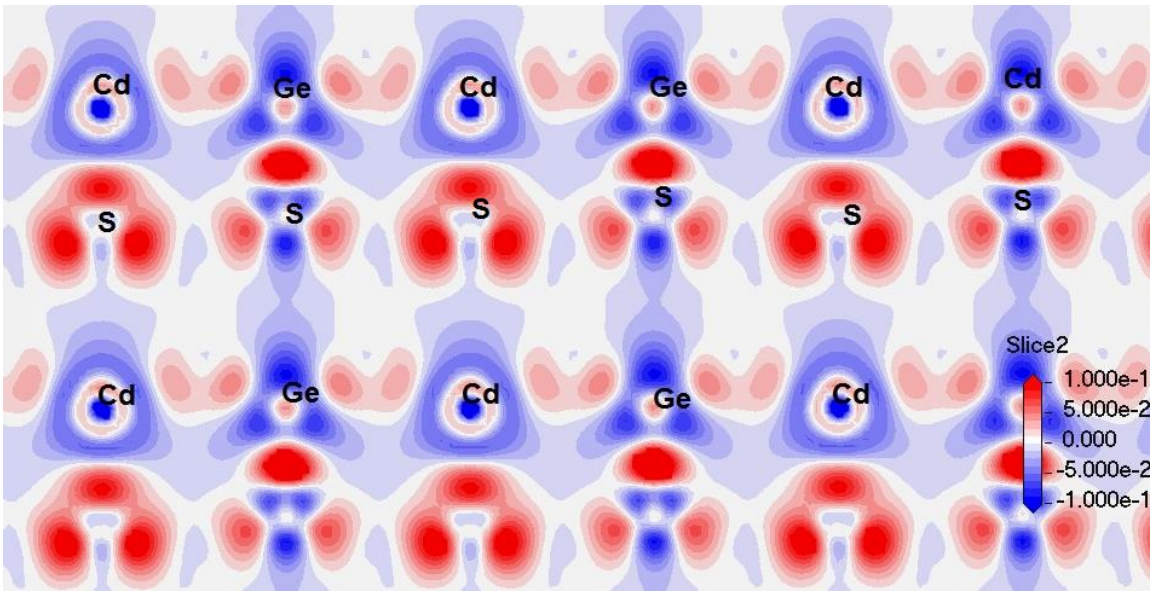

Fig. 8. Calculated cross-section of the electron density difference in the plane parallel to the (*a*, *b*) (top) and (*a*, *c*) crystallographic axes (bottom) for $Cu_2CdGeS_4$. The scale is in electrons/Å$^3$.

One can clearly see that following the space charge transfer the Ge-S and Cd-S bonds are more covalent than the Cu-S bonds, which are more ionic, since the Cu electron density is close to spherical, like that one of free ions. The effective Mulliken charges of all ions are given in Table 2; as can be seen from the Table, the effective charges of all ions are very different from those formally expected from the chemical formula.

Table 2. Effective Mullikan populations and charges (GGA/LDA) for $Cu_2CdGeS_4$.

|    | s | p | d | Total | Charge |
|----|-----|-----|-----|-------|--------|
| Cu | 0.60/0.60 | 0.69/0.80 | 9.80/9.79 | 11.09/11.19 | -0.09/-0.19 |
| Cd | 0.66/0.72 | 0.79/0.87 | 9.98/9.97 | 11.43/11.56 | 0.57/0.44 |
| Ge | 0.88/0.77 | 2.04/2.12 | 0.00/0.00 | 2.92/2.89 | 1.08/1.11 |
| S  | 1.83/1.81 | 4.53/4.49 | 0.00/0.00 | 6.36/6.30 | -0.36/-0.30 |

**6. Modeling of pressure effects on the structural, electronic and optical properties**

Application of the hydrostatic pressure to solids can significantly alter their properties. Additionally it gives useful information concerning the voids, which also may change degree of optical anisotropy. The pressure effects on the structural and electronic properties of $Cu_2CdGeS_4$ were modeled by optimizing the crystal structure and calculating the electronic and optical properties at the elevated hydrostatic pressure, in the range from 0 to 20 GPa with a step of 5 GPa; such an option is implemented in CASTEP and allows for finding an optimized structure at any axial or hydrostatic pressure. Table 3 collects the structural parameters and band gap values for $Cu_2CdGeS_4$ (the latter are given without the above-given scissor operators for the GGA and LDA calculations) obtained as the results of the performed calculations for elevated pressures.

Table 3. Calculated band gaps (in eV) and lattice parameters *a*, *b*, *c* (all in Å) for $Cu_2CdGeS_4$ at different pressures.

| Pressure, GPa | GGA | | | | LDA | | | |
|---|---|---|---|---|---|---|---|---|
| | Band gap | *a* | *B* | *C* | Band gap | *a* | *b* | *c* |
| 0 | 0.372 | 7.77051 | 6.60668 | 6.34993 | 0.682 | 7.52811 | 6.39153 | 6.16532 |
| 5 | 0.638 | 7.61325 | 6.44695 | 6.23318 | 0.909 | 7.40823 | 6.26678 | 6.07917 |
| 10 | 0.860 | 7.49579 | 6.31722 | 6.14833 | 1.096 | 7.31123 | 6.16213 | 6.01418 |
| 15 | 1.050 | 7.39973 | 6.20201 | 6.08800 | 1.253 | 7.23251 | 6.06543 | 5.96445 |
| 20 | 1.212 | 7.33297 | 6.08602 | 6.03987 | 1.382 | 7.17622 | 5.96227 | 5.92673 |

The data from Table 3 are visualized in Figs. 9 - 11. Since the band gap of the studied crystal is a direct one, it is increasing with pressure, and its dependence on pressure is well described by a linear function with the slopes of about 0.035 eV/GPa (LDA) and 0.042 eV/GPa (GGA), as Fig. 9 shows.

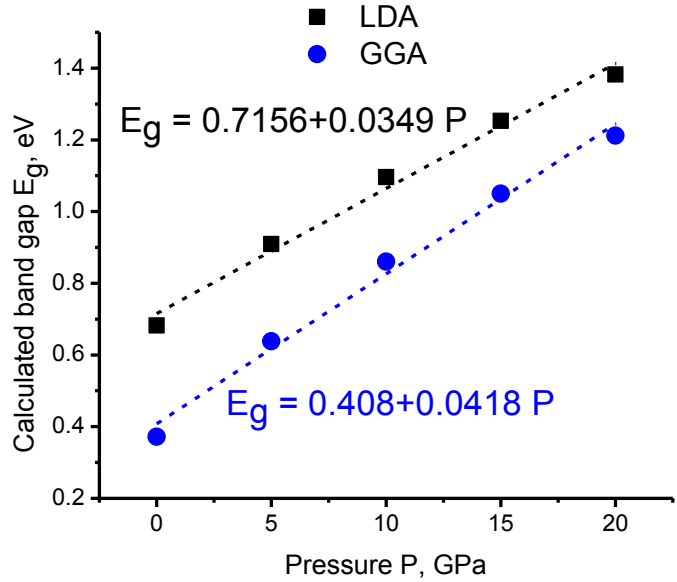

Fig. 9. Calculated band gap (symbols) and a linear fit (dashed lines) as function of pressure *P* for $Cu_2CdGeS_4$.

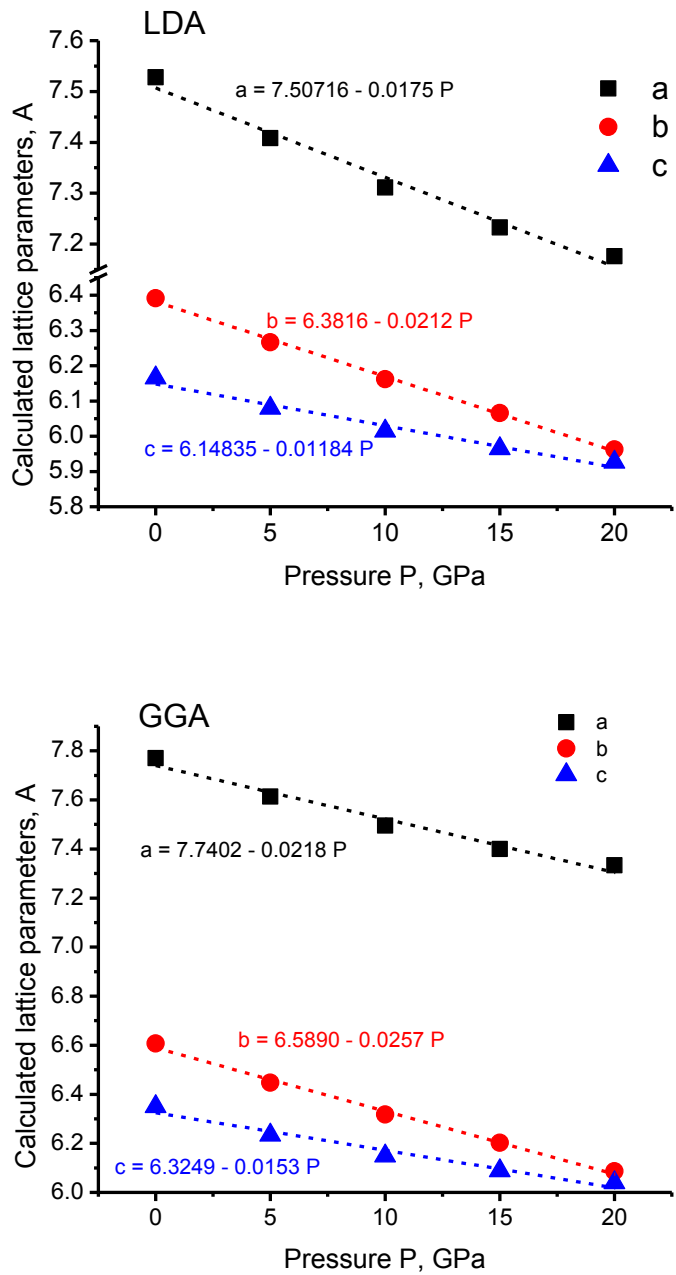

Fig. 10. Calculated lattice parameters (symbols) and their linear fits (dashed lines) as function of pressure $P$ for $Cu_2CdGeS_4$.

Dependencies of the calculated lattice parameters on pressure are shown in Fig. 10. All three parameters *a*, *b*, and *c* decrease linearly with increasing pressure, with the largest (in absolute value) pressure coefficient for the *b* parameter in both GGA and LDA calculations, which means that the largest compressibility of $Cu_2CdGeS_4$ is realized along the $b$ crystallographic axis.

Finally, after the lattice parameters at various values of hydrostatic pressure are found, the volume $V$ of a unit cell can be readily calculated, and its dependence on pressure can be fitted to the Murnaghan equation of state [53]:

$$\frac{V}{V_0} = \left(1 + \frac{PB'}{B}\right)^{-1/B'}, \qquad (3)$$

where $V_0$ denotes the unit cell volume at the ambient pressure, $B$ is the bulk modulus, and $B' = dB/dP$ is its pressure derivative.

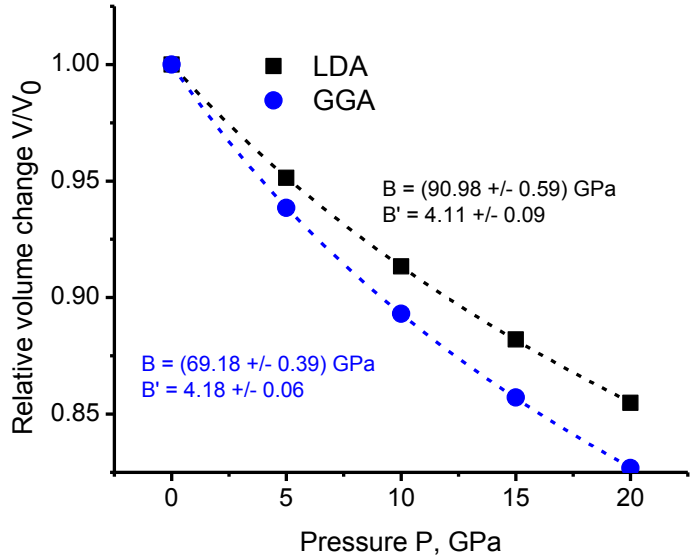

Fig. 11. Calculated dependence of the relative volume change $V/V_0$ (symbols) and fits to the Murnaghan equation of state (dashed lines) as function of pressure $P$ for $Cu_2CdGeS_4$.

Results of application of Eq. (3) to the calculated variation of the $Cu_2CdGeS_4$ unit cell volume are shown in Fig. 11. Estimations of the bulk modulus give the values of about 91 GPa (LDA) and 69 GPa (GGA). To the best of our knowledge, these are the first theoretical estimates of the bulk modulus for $Cu_2CdGeS_4$.

The obtained values of the bulk modulus of $Cu_2CdGeS_4$ are in favorable agreement with the bulk moduli of other chalcogenides possessing sulfur anions, such as LuS (108

GPa [54]), YS (95.86 GPa, [55]), PbS (56 - 66 GPa, [56]). Because the above-mentioned crystals are cubic and there is no optical anisotropy for them, one can conclude that the voids and defects do not play a crucial role in formation of the elastic properties of $Cu_2CdGeS_4$, in spite of their importance for the optical properties of this crystal. Following this reason the titled crystals may be promising for various kinds of optoelectronic waveguides [57].

## 7. Conclusions

The present paper reports the results of the combined experimental photoinduced changes in quaternary semiconducting $Cu_2CdGeS_4$ chalcogenide single crystals studied using the cw 532 nm green laser with energy density about 0.4 J/cm$^2$ together with the DFT analysis of its electronic structure and optical properties. The spectral maximum of the photoinduced anisotropy was observed at spectral energies equal to about 1.4 eV (energy gap equal to about 1.85 eV) corresponding to maximal density of the intrinsic defect levels. The absorption edge of our samples varied from 1.83 eV to 1.86 eV depending on the probing light polarization (or non-polarized light). This difference in energy positions of principal spectral maxima correlates well with the anisotropy of absorption edges determined by difference in the space charge density distribution for the Cu-S-Cu bonds formed mainly by the Cu 3d and S 3p states and the Cd-S-Cd bonds. As a consequence the coordination of the S anions is different due to different degree of the hybridization of the S 3p states with the Cu and Cd states.

The type of the absorption edge corresponds likely to the direct optical transitions; however there is some obvious long-range tail responsible for the presence of the defect states in the band energy gap. It is crucial that the photoinduced changes of absorption is higher for the unpolarized light with respect to the polarized states which may highlight the principal role of the local intrinsic defect levels Such behavior may reflect a fact that presence of the disordered state may be crucial here because these states are not sensitive to the optical transition rules. The calculated optical reflection spectra show two principal anisotropic spectral maxima. The LDA-calculated optical spectra are more anisotropic with respect to the GGA-calculated ones. The origin of such anisotropy is confirmed by the anisotropy of the charge density distribution for the two principal crystal planes. The cross-section of the electron density difference in the space between the atoms in the $Cu_2CdGeS_4$ crystal lattice indicate presence of two types of chemical bonds: ionic and covalent, with the former being more characteristic of the Cu-S bonds and the latter being more characteristic of the Ge-S and Cd-S bonds. Different cross-sections show the anisotropy of the bond lengths in different directions. Particularly important is the anisotropy for the Cu and Ge atoms forming the principal band energy gaps. The obtained experimental data show that the role of the anisotropy is not so crucial for the photoinduced changes like the presence of the randomly oriented intrinsic defect states. The results of the benchmark first-principles calculations for the structural, electronic, optical and elastic properties of the title compound are also given. According to these calculations, $Cu_2CdGeS_4$ is a direct band gap semiconductor. Peculiar features of the optical anisotropy correlate well with the electronic band structure and electron density distribution. The pressure coefficient for the direct band gap was estimated to be about 0.03 - 0.04 eV/GPa; dependence of the band gap and lattice parameters on the applied hydrostatic pressure are all well described by the linear functions. From the dependence of the lattice volume on the pressure, the bulk modulus was evaluated as 91 GPa (LDA) and 69 GPa (GGA). From the performed calculations and comparison with the results obtained for other chalcogenides, the voids and related defects do not manifest themselves in the elastic properties of $Cu_2CdGeS_4$.

**Acknowledgments**

M.G. Brik appreciates the financial support from i) European Social Fund's Doctoral Studies and Internationalisation Programme DoRa, ii) European Union through the European Regional Development Fund (Center of Excellence ''Mesosystems: Theory and Applications'', TK114) and iii) Marie Curie Initial Training Network LUMINET, grant agreement No. 316906. Dr. G.A. Kumar (University of Texas at San Antonio) is thanked for allowing to use the Materials Studio package.